# Optical Kerr soliton microcombs for high bandwidth communications

Bill Corcoran[1,6], Arnan Mitchell[2,6], Roberto Morandotti[3], Leif K. Oxenlowe[4], and David J. Moss[5,6*]

[1] Photonic Communications Laboratory, Dept. Electrical and Computer Systems Engineering, Monash University, Clayton, VIC 3800, Australia

[2] School of Engineering, RMIT University, Melbourne, VIC 3001, Australia

[3] INRS –Énergie, Matériaux et Télécommunications, Varennes, QC J3X 1S2, Canada

[4] DTU Fotonik, Technical University of Denmark, Lyngby, Denmark.

[5] Optical Sciences Centre, Swinburne University of Technology, Hawthorn, VIC 3122, Australia

[6] ARC Centre of Excellence in Optical Microcombs for Breakthrough Science (COMBS), Australia

*E-mail: dmoss@swin.edu.au

**Abstract**

Micro-combs - optical frequency combs generated by nonlinear integrated micro-cavity resonators – have the potential to offer the full capability of their benchtop comb-based counterparts, but in an integrated footprint. They have enabled breakthroughs in spectroscopy, microwave photonics, frequency synthesis, optical ranging, quantum state generation and manipulation, metrology, optical neuromorphic processing and more. One of their most promising applications has been optical fibre communications where they have formed the basis for massively parallel ultrahigh capacity multiplexed data transmission. Innovative approaches have been used in recent years to phase-lock, or mode-lock different types of micro-combs, from dissipative Kerr solitons to dark solitons, soliton crystals and others. This has enabled their use as sources for optical communications including advanced coherent modulation format systems that have achieved ultrahigh data capacity bit rates breaking the petabit/s barrier. Here, we review this new and exciting field, chronicling the progress while highlighting the challenges and opportunities.

## I. Introduction

The world's optical fibre communications network forms the backbone of the global internet and now supports hundreds of terabits of data every second, with demand growing exponentially at about 25% per annum [1]. The success of optical data communications is a result of the extremely wide optical bandwidth available in optical fiber. This, when combined with erbium doped fibre amplifiers (EDFAs) has enabled systems with enormous data capacity, able to transmit, modulate, amplify and switch at rates unachievable with other technologies [2]. Commercial long-haul systems now operate in the wavelength range of 1530-1610nm (the telecom C and L bands), representing over 10 THz of optical frequency bandwidth in the near infrared, a value that new technologies are continually expanding. With the latest electronic chip interface speeds limited to around 100GHz, the spectrum must be sliced into multiple independent channels in order to fully exploit the available optical bandwidths. This technology underpinned the rapid expansion of the global internet in the 1990s, termed wavelength division multiplexing (WDM), and has evolved to extremely sophisticated ultrahigh capacity transmission links that employ massively parallel WDM systems using highly advanced coherent modulation formats [2]. These systems typically use independent discrete laser sources for each channel carrier, with data modulated onto these carriers using external modulators. In this way, each individual channel can interface directly with state-of-the-art electronic interfaces, making the entire ultra-high bandwidth optical communications spectrum available for digital communication technologies. In this approach, spectral gaps, or "guard-bands" must be left to allow for laser frequency drift to avoid interference between neighbouring wavelength channels, and this reduces the spectral efficiency. The continually increasing requirement for higher data rates in fibre optic communications systems have placed extreme demands on optical components and subsystems, particularly for transmitters and receivers.

Optical frequency combs (OFCs) [3-5] have emerged as attractive sources for communications [6, 7] for many reasons, including reducing or eliminating the need for guard-bands, thus freeing up much more of the optical spectrum to be used for data transmission. Recently, integrated OFC sources, termed "Micro-combs" or "Kerr combs", have emerged as a powerful and innovative approach to help advance massively parallel WDM systems. Based on nonlinear integrated micro-cavity resonators, [8-11] they now generate 100's of wavelengths coherently on a single chip. Initially discovered in 2007 [8] they have been highly successful as integrated sources for many applications [12], particularly for ultrahigh capacity massively parallel multiplexed data transmission [13-18] aided by their realization in integrated mass-manufacture compatible (CMOS: complementary metal oxide semiconductor) form [19-21]. Achieving coherent states, termed dissipative Kerr solitons (DKS), [22] opened the door to many novel temporal soliton coherent microcomb states, including soliton crystals, [23, 24] dark solitons [25], laser cavity solitons, [26] and others, [27] each having their own unique advantages.

This paper reviews ultra-high bandwidth optical data communications based on integrated optical Kerr microcombs. We discuss promising new forms of microcombs, not yet applied to communications but which may offer significant advantages. Finally, prospects and paths to achieve even higher levels of performance are outlined, followed by speculation on which optical communications markets microcombs may have the first and potentially lasting impact on.

## II. Optical Microcombs

Optical microcombs have become a huge field of research, with the promise of offering the full capability of OFCs [3-5] on an integrated chip. Many types of microcombs have been reported [9-11] including DKS states, [22] dark solitons [25] and soliton crystals. [23]. Figure 1 chronicles the key milestones in microcombs, beginning with the first Kerr microcomb in 2007 [8] soon followed by CMOS compatible integrated microcombs. [19, 20] These initial microcombs featured what later became known as simple primary combs or Turing patterns [9, 10] that were already highly stable. However, they featured a wavelength spacing, determined by the peak modulational instability gain, that was too wide (100's of GHz to THz) for most applications. Microcombs were soon able to achieve spacings at the resonator FSR [15 but the highly complex dynamics and nonlinear physics were not understood well enough to enable modelocking, resulting in only partially coherent or even chaotic combs. Subsequent investigations [28, 29] led to the realization that microcombs can be described by the well-known Lugiato-Levevre equation, [9, 10, 30] and this in turn led to the ability to phase or mode-lock the comb lines, yielding temporal solitons that were named DKS states. [22] This evolved into feedback-stabilized DKS combs [13], thus improving their coherence, noise and stability further. Subsequent advances achieved ultralow power operation [31], the ease and reliability of comb generation through control loops [32], turnkey generation through the device physics [33, 34], self-starting robust operation [35, 36], high conversion efficiency, [36, 37] wide bandwidths [38], and comb spacings down to 10's of GHz for microwave applications [39,40]. In parallel with this, increased levels of integration have been achieved, including heterogenous integration of the microcomb platforms such as SiN with SOI [41], III-V devices [42], and lithium niobate [43].

DKS states are 'bright pulse' states with higher peak power than the background pump light [22]. Bright soliton states display robustness against small environmental perturbations, and when implemented inside a fiber loop, can exhibit long-term stability, self-starting and self-recovery [32]. A key performance metric for microcombs is conversion efficiency of the incoming pump seed light to comb lines. Although DKS states have been one of the most successful microcombs because of their high coherence, number of comb lines and spectral flatness, a challenge has been that the conversion efficiency is limited to about 4%. This arises because the Lugiato-Lefevre equations [9, 10] require a CW background, causing single solitons to contain a relatively low fraction of the optical energy in the cavity. This yields a high residual pump power and relatively low power per-comb-line. Together with the relatively complex pump tuning dynamics needed to initiate DKS states, this has motivated the search for new microcombs, also to drive up the power per comb line, increasing the OSNR and conversion efficiency.

A complementary approach is that of dark pulse states, or 'dark solitons', [25] which in some sense are "inverse" solitons – a negative soliton state imprinted on a high CW background. Dark pulses (DPs) can occur when the balance between nonlinearity and dispersion is inverted compared to micro-resonators that support DKS states. While DKS microcombs require anomalous, or positive chromatic dispersion (i.e. the dispersion coefficient $D > 0$ ps/(nm.km)), in most materials where the Kerr nonlinearity ($n_2$) is positive, dark pulse states need normal dispersion, ($D < 0$). This has been an attractive feature of DP states since achieving anomalous dispersion in silicon nitride nanowires requires growing relatively thick (>700nm) films in which intrinsic film stress tends to lead to film cracking, [44] although to a large degree this has been solved [45-47]. A key feature of DP states is that their conversion efficiency is significantly increased compared to DKS states, to between 30 - 40 % [25], mainly because their intracavity energy is much higher, thus increasing the fraction of energy that resides in the comb lines. However, generating the dark-pulse Kerr combs in single ring cavities with normal dispersion requires a delicate linear coupling between transverse modes of the ring waveguide, requiring active feedback to stabilise the comb output [25]. While feedback systems are commonplace and can be co-packaged with microcombs, their use still adds to complexity, cost, footprint and energy consumption. DP states typically feature relatively broad spectra and can be generated using a wide range of microring resonator designs as long as they display normal dispersion. Coupled rings with normal dispersion can generate dissipative photonic molecule solitons, which (in contrast to single ring DPs) are thermally self-stabilised, removing the need for active feedback. These microcombs can be turn-key operated, have conversion efficiencies of about 50%. [48, 49, 50, 51, 52], and can have spacings of 50 GHz [52]. Moreover, the spectral profile of these combs is relatively flat and wide and constrained to a bandwidth of about 60 nm (at 20 dB down). This means that comb lines can be generated with relatively consistent shapes, and high power over a desired targeted spectrum, as opposed to DKS states where the comb profile is limited to a very broad sech-shape [10,13], resulting in a significant fraction of the comb lying outside the telecom bands.

Soliton crystals (SCs) were first discovered [23] in microdisk resonators and yielded a rich variety of comb states, followed by their demonstration in CMOS compatible microcombs having only a few spatial modes. [53] These enabled many demonstrations of microwave photonics [24, 54-57] and other applications including ultrahigh bandwidth optical neuromorphic processors.[58] Whereas both DKS and DP states typically feature a single pulse circulating around a micro-resonator, soliton crystals consist of multiple soliton states, providing multi-pulse microcombs with many lines. The advantages of SCs include a relatively high conversion efficiency of up to 75% [23], as well as stability and ease of generation. [14] These arise because SCs lack a significant 'soliton step' – a discontinuous drop in power taking place during the transition from the chaotic state to the soliton state - because the in-cavity power of SCs is very similar to that of the chaotic state. This means that there is very little thermo-optic or nonlinear optical (eg., Kerr) induced shift in the resonator wavelength during the transition, which results in simple and robust – close to turn-key – operation. Indeed, SCs can be generated by slowly – even manually – tuning the pump wavelength [14]. As with bright soliton states, soliton crystals are also resilient to small environmental perturbations and do not require active feedback to maintain the state. SCs can be "prefect" (PSCs - e.g. [59]), or include "defects" in their structure (e.g. [23]). PSCs are multiple evenly-spaced solitons, which often yield an undesirably large comb line spacing for communications applications. Defect state SCs produce a comb where the line spacing matches the resonator FSR, which is often desirable for communications. However, these SCs produce a spectrum that is not flat but has a "scalloped" shape, featuring many possible states. This can be both an advantage in terms of versatility, but also a disadvantage in that some effort can be required to produce the same SC state [60]. The non-flat spectral profile means that some comb lines have relatively low power compared to the pump, despite the high generation efficiency. Also, SCs require engineered dispersion mode crossings, and while rings with the required crossings can be reproducibly realized through careful device engineering [60], it nonetheless complicates the design process.

Important for all types of microcombs are high conversion efficiency, stability, bandwidth, comb line powers, and reduced pump power and FSR. [13, 14, 18] Conversion efficiency, defined as the ratio of the comb power residing in the comb lines relative to the power in the pump, is critical. For a given pump power, a lower conversion efficiency reduces the per-line comb power and hence the OSNR. A high OSNR is key to achieving high spectral efficiency - both at the transmitter and receiver sides. Another issue that is often overlooked is that, while generally having a broad-band and flat optical comb spectrum is beneficial, there is an optimum width. Very broad-band combs having a substantial fraction of the comb energy lying outside the wavelength bands of interest result in wasted energy. However, microcombs have generally demonstrated high power efficiency, resulting in OSNRs of 30 - 40 dB when amplified to powers that would reach watt-level for a full comb [13,14,16,18,61] – ideal for data modulation and transmission. All these advantages, together with compactness and cost-effectiveness, has led over the last decade to significant progress in using microcombs for high-capacity optical communication systems.

## III. Microcombs for Optical Communications

OFCs in general have been used for ultrahigh bandwidth communications because of their short pulsewidths [62, 63], or coherence and stable frequency spacing for optical super-channels [64-66]. Figure 2 shows the timelines of state-of-the-art communications systems for both benchtop OFCs and microcomb systems. Early benchtop OFC systems based on active fibre mode-locked lasers employed legacy modulation formats [62,63], and while they did achieve significant coherence, they suffered from phase discontinuities caused by supermode noise. Multiple approaches have been used to both reduce the phase noise (e.g. [64-66]) and size (e.g. [67-71]) of mode-locked lasers or comb systems based on these, generally resulting in a tension between size, performance and device complexity. This has meant that ultrahigh-capacity communication demonstrations with OFCs have often used benchtop-based OFCs [6,7,64].

Optical microcombs are then a highly attractive in comparison, in principle only requiring a pump source and microresonator (Table 1). They are potentially coherent light sources with precise comb spacings, similar to benchtop combs but in an ultrasmall footprint. In the last decade microcombs have been extremely successful for optical communications. Even early microcombs were successful at massively parallel optical communications, [15] achieving transmission rates of 2 Tb/s, despite their semi-coherence. This highlighted (Figure 2) that microcombs could allow for very low guard bands, supporting the well-known approach of optical "superchannels" in order to increase spectral efficiency. Achieving a high spectral efficiency is critical since it determines the fundamental limit of data-carrying capacity for a given optical communications bandwidth. [1, 2]

Comb coherence is a key property for many applications but particularly for data transmission employing ultrahigh coherent modulation formats [1, 2]. With the successful modelocking of microcombs with DKS states, [22] massively parallel optical communications at 30.1 Tb/s with a single microcomb using the full C and L telecommunication bands and a coherent modulation format of 16-QAM was demonstrated (Figure 3). [13] This work also used two separate microcombs multiplexed together to address the wide comb spacing (100 GHz) compared to modulation interface rates of ~50 GHz. This resulted in a further increase in both the data rate to 50.2 Tb/s and spectral efficiency to 5.2 b/s/Hz. [13] This was enabled by the stability of the two combs as well as the flat optical spectrum with an optimum 3-dB width of 6 THz. A relatively high per comb line power of >-20 dBm allowed a comb line OSNR of 40 dB after amplification to powers compatible with coherent modulation. Although after modulation and re-amplification (compensating for modulator loss) the OSNR was reduced to 23 dB, theoretical models [18, 73], have shown that a higher OSNR could be achieved using a narrow filter before the modulator. There are several techniques that may be used to achieve the same aim [74].

An even higher order modulation format of 64-QAM was subsequently reported [16] (Figure 4) using dark pulse microcomb solitons having both higher conversion efficiency and an OSNR comparable to both DKS and benchtop combs, at > 35 dB. Notably, the power consumption was reduced to 100's of mW from watts, while the conversion efficiency improved to over 20%. With the ability to support 64-QAM, this work points the way to significantly improving the total power efficiency (bits/second/Watt, or bits/joule) of a comb-based superchannel. The wide line spacing of 230 GHz, however, limited the comb to 20 lines across the C-band, restricting the spectral efficiency to 1.1 b/s/Hz and data rate to 4.4 Tb/s. Simulations of the dark pulse system performance, however, pointed the way to achieving full C-band coverage at a very high OSNR with a dark pulse microcomb systems by adjusting parameters.

At about the same time, ultrahigh data rates were demonstrated with a single soliton crystal (SC) microcomb [74] (Figure 4) with 80 comb lines and an OSNR > 30 dB OSNR. The comb had a bandwidth of 4 THz and spacing of 49 GHz – very close to the telecommunications standard of 50Hz. By modulating a single sub-band onto each comb lines, a high spectral efficiency of 8.2 b/s/Hz was achieved. Although slightly lower than previous microcomb records [13] because the higher baud rate increased the system transceiver noise, the larger number of high OSNR comb lines over a greater bandwidth enabled a considerably higher overall data rate at 32.5 Tb/s, similar to results with a single 100 GHz device covering the full C+L bands [13]. The capacity of the single SC microcomb system was subsequently [14] increased using virtual sub-banding where, instead of modulating each carrier with a unique modulator, single sideband modulation enabled two sub-bands to be independently modulated onto a single comb line. Modulating the signal at 64-QAM with two 23 Gbaud sub-bands per carrier, a spectral utilization of 94% was achieved, yielding a net data rate of 39 Tb/s and spectral efficiency of 10.1 b/s/Hz. These represented a new record for a single microcomb over standard single mode fiber. Importantly, in both demonstrations the SCs were easily generated and stablized using open-loop control (no active feedback), providing a simple and robust microcomb. This, together with the high transmission capacity and spectral efficiency were a direct result of the very high conversion efficiency between the injected CW wave and the soliton crystal state of about 40%. Lastly, these demonstrations were performed over a field-installed fibre network link in Melbourne's metropolitan area, highlighting that microcomb optical communication systems are compatible with real-world fibres and with the associated environmental effects.

A key challenge for microcomb-based communications has been to lower the comb spacing to better match the bandwidth of available electro-optic interfaces. In this respect, coherent Kerr combs based on larger sized micro-disk resonators or large integrated microresonators offer the advantage of exhibiting smaller comb line spacings of 22.1 GHz line [61], allowing a high modulation format of 64-QAM at 21.5 Gbd (Figure 4), and resulting in a very high spectral efficiency optical superchannel [61] of 97%, or 10.1 b/s/Hz. This demonstration used only 52 lines from a DKS state over a bandwidth of only 1.1 THz, jointly enabled by the low comb spacing (4.5x lower than [13]) and low 150 mW pump power. However, loss in the microcomb package from free-space tapered fiber coupling produced per-line comb powers < -30 dBm (1 μW). Importantly, though, this work represented a key milestone showing that microcombs could support superchannels at high spectral efficiencies, comparable to independent laser sources. This highlighted the potential for microcombs to replace individual lasers or benchtop combs for optical communications. The challenge is to increase the per-comb line power and hence the per-comb line OSNR at typical operating powers.

To go beyond 10's of Tb/s towards Pb/s, spatial division multiplexing (SDM) of independent signals not only in wavelength, but in space [1, 2, 6, 7, 18, 64] is required using independent or multi-core fibres, multi-mode fibres, or combinations of these. This technique enables comb lines from a single microcomb source to be split into multiple independent data streams, dramatically increasing the overall bit rate. The trade-off is that the power of every comb line in each fiber core is reduced, reducing the OSNR after amplification, ultimately limiting the data capacity. Recently, [18] SDM was employed over a 37-core, 7.9-km-long fibre, using 223 wavelength channels from a single microcomb source to break the Pb/s barrier for the first time, at 1.84 Pbit/s (Figure 4). This was accomplished using a stabilized dark-pulse soliton Kerr frequency comb with a 105 GHz line spacing and conversion efficiency of 13%. Here, the comb lines were modulated to generate a sub-comb with a smaller spacing of 35 GHz, before encoding the data onto the comb lines. By covering 7.8 THz across both the C and L bands, the microcomb supported close to 50 Tb/s per fibre core – comparable to previous results [13] achieved with two combs, and 20% higher than the SC demonstration [14], but importantly using 37 spatial modes. The trade-off is that the comb power is shared with every SDM channel over 37 fibre cores and so the individual OSNR in each SDM channel decreased [18] limiting the spectral efficiency to 6.4 b/(s.Hz).

For all these demonstrations, the advantages of microcombs compared to discrete lasers include eliminating guard bands and efficiently providing many optical carriers, while still achieving high phase stability and high OSNRs. System designs assume that the optical carriers contribute negligible noise, and while this was true for discrete lasers, it was not for OFCs and early microcombs. For a given overall power, increasing the optical bandwidth or number of wavelengths [75] reduces the single line power, limiting its OSNR. Increasing the per comb line OSNR, optical bandwidths and number of wavelengths all require increasing the overall system power, which is contrary to the spirit of low power systems. Phase stability is largely determined by the intrinsic linewidth of each comb line and the stability of the comb itself – ie., the ability to operate without large, sudden shifts in phase. The intrinsic linewidth is determined by the active cavity of a pulsed laser system [26,35], or by the seed laser used as a basis for optical frequency comb generation [10,11,23,25], both of which can achieve high phase stability owing to their small footprint.

For the receiver side where comb lines are used as local oscillators, the single line power of microcombs may still be too small. Likewise, at the transmitter side, lowering the power per comb-line lowers the modulated signal launch power [75]. Of course, one can amplify an optical frequency comb to increase the per comb-line power *before* modulation at the transmitter side or before use as a local oscillator at the receiver side. This, however, has ramifications on the system OSNR [13, 18, 76] and can be modelled as an amplitude-dependent error vector addition, limiting achievable system data rates [72,73]. These trade-offs are compounded if the comb spectrum is not spectrally flat. Overall, OSNR is one area where there is still room for improvement.

We note that the spectral efficiency achieved by microcomb systems [14, 18, 61] with OSNRs close to 30 dB was comparable to the best achieved with benchtop combs [6,7]. Modelling of microcomb based systems indicates that OSNRs of 30-40 dB are required and this has been achieved experimentally [13,14,16,18,61]. The achievable single comb line power at the output of a microcomb is limits the OSNR after amplification. Hence, higher output power microcombs support higher OSNRs with larger numbers of channels, enabling even higher data rates. So, there is clear scope for not just generating broad combs at low power, but in tailoring comb generation to provide high power lines within a desired optical bandwidth.

However, with unbound parallelism in space through massive SDM, 100 Pbit/s data transmission using a single comb source is theoretically attainable with current microcombs [13, 18]. This raises the question if there is an optimised usage of microcombs for SDM systems, e.g., is it better to have high spectral efficiency per core or to have a massive total capacity through many cores with reduced spectral efficiency in each core, essentially balancing spectral efficiency with energy efficiency of the transmitter. Regardless, SDM based microcombs have been demonstrated [18] as having strong potential for not only massively parallel multi-wavelength sources, but to support massively parallel communications across multiple SDM modes.

Summarizing, we note that all of these types of microcombs each have their advantages and disadvantages for communications - there is no single winner. All have achieved remarkable success at realizing world record data transmission. Nonetheless, the advances in improving microcomb performance has continued unabated and this has already yielded new forms of soliton microcombs that are promising for future optical communications.

In many demonstrations of optical communications in systems using OFCs, the emphasis has been on replacing banks of lasers for increased frequency stability and power efficiency. However, the phase and frequency locking of comb lines can enable much more. The signal digital processing needed to compensate for laser frequency drift and phase noise can be reduced by realising that these parameters are correlated on WDM channels derived from an OFC, resulting in architectures that should enable power savings [77]. Digital pulse shaping, generally used to avoid WDM channel overlap and interference in high spectral efficiency systems, can be relaxed (maybe even removed) by joint detection of channels supported by OFCs [78], which may enable better use of the limited dynamic range of digital to analogue converters at the transmitter side. Multiple independent WDM channels can be generated from a frequency comb by taking advantage of a novel time-lens modulation architecture, providing a novel path to reducing the electro-optic hardware requirements in comb-based transmitters [79]. Using phase correlations on multiple lines enables the outputs of independent modulators to be coherently combined to generate and measure a single channel beyond the bandwidth of electro-optic interfaces [80,81].

Each of these demonstrations point to ways that microcombs can not only provide a compact source of multiwavelength light, but also a method to help drive efficiencies in the rest of the components in an optical communications system. While past demonstrations exploring the use of microcombs for optical communications have driven great interest, it would seem that we are just beginning to really understand their future impact.

## IV. Advances in Microcombs

New types of soliton microcombs include laser cavity solitons (LCS) [26,35,36], and high-efficiency bright states using couple resonator "photonic molecules" [82], and while they have not yet been used for communications they represent potentially attractive approaches. As outlined earlier, photonic molecule micro-combs have been demonstrated in normal dispersion coupled micro-resonators [48,49], overcome issues of comb stability experienced by dark pulse solitons. Recently, this work has been extended to rings operating in the anomalous dispersion regime, with up to 54% conversion efficiency [82]. While the required fabrication of two coupled resonators with the necessary characteristics may increase fabrication complexity, the flexibility in defining combs with these molecules may help to overcome many drawbacks. Moreover, photonic molecule microcombs appear to have deterministic generation, indicating the potential for turn-key operation, with the potential to be coherent with low noise, well suited to communications. One trade-off is that the highest efficiencies for this approach come with reduced the pump power, and so tracking the per line power can be important for communications systems. Current demonstrations [82] have also needed a higher power to initiate the soliton state than to run it, and so a microcomb system design based on this approach needs to enable this short-term higher power operation.

Microrings nested in an amplified fibre-loop cavity have been used to generate "laser cavity-solitons" (LCS), [27, 35, 36] which are distinct from other microcavity based solitons. They comprise a composite nested cavity somewhat resembling a miniaturised fibre mode-locked laser, but with the micro-resonator providing parametric mixing and frequency mode definition, and a broadband gain source to generate and sustain the comb. By using a short gain medium, a compact amplified fiber loop can be obtained so that only a single resonance of the loop is contained within a micro-resonator resonance. This results in mode-locked cavity soliton pulses with low linewidth that are super-mode noise free. Most importantly, LCS states can be made fully self-starting, displaying naturally self-emergent behaviour, [35] essential for turnkey operation. This is analogous to passive modelocking achieved in the 1960s for modelocked lasers [83]. In [35], a 48.9 GHz FSR microcomb was demonstrated with a 20-dB bandwidth of 20 nm, covering a large part of the C-band. Importantly, they can be set to operate in any desired state – eg., single, dual solitons etc., simply by setting the maximum amplifier gain and adjusting the in-loop free space control. Just as important as turnkey operation, they are robust to disruption – naturally recovering to the original operation state if disrupted. Finally, they represent the most efficient microcomb source reported to date, experimentally achieving >95% efficiency with 100% theoretically possible, resulting from the fact that they do not require a CW background like the Lugiato-Lefevre based solitons (all types, including photonic molecules [82]), thus freeing up all of the optical energy to reside in the comb lines. While LCS systems exploit both free-space and fibre components, achieving full integration with phase controls and wideband amplification using either semiconductor optical amplifiers or on-chip erbium amplifiers [84] is possible. Further, the bandwidth of LCSs is not fundamentally limited – only by the filtering and amplifier components. By broadening the spectra together with all the other advantages, laser cavity-solitons are very promising for field-ready prototypes for commercial use.

Finally, there has been significant progress in integrated all microcomb components onto a single chip, [34] including integrating a SOA to form a tunable external cavity pump laser with one or more micro-resonators. [31] By tuning on-chip microheaters, a DKS state was initiated and stabilized, requiring very low pump power (<100 mW). In [31] a semiconductor laser diode was coupled directly to a micro-resonator, which used injection locking between the laser and back-reflected pump light from the resonator, and enabled self-starting turnkey operation. Interestingly, that DKS state did not require active fast laser tuning but instead used a simple ramp in current on a millisecond time-scale together with the frequency locking properties of optical injection locking. So far, however, these integrated approaches have yielded fairly low per-comb-line power, as they require amplification for communications applications, thus reducing their OSNR. Emerging techniques for stabilizing and tuning microcombs all optically though may offer advantages in reducing or eliminating external electronics or other feedback control methods [85-89].

## IV. Future perspectives

While there has been tremendous progress in massively parallel ultrahigh bandwidth optical communications based on microcombs, the question remains of exactly when and where they might have real-world impact (Figure 5). Because of their compact nature, their potential scope of applications is truly diverse – from ultrashort optical interconnects and board-board communications, to datacentres and long haul communications. The requirements vary dramatically between these different network scales as does the comparative advantages of microcombs.

For short reach applications, SWaP (size, weight and power) and particularly cost are paramount, more so than overall performance. Hence microcombs are attractive since they are intrinsically integrated, and have been co-integrated with silicon and III-V semiconductors. [41, 42] On a larger scale, intra-data centre compatible communications with microcombs have also been demonstrated. [90] The competition for microcombs is arguably benchtop frequency combs that have comparable per-line power to discrete lasers. This is important since it determines the carrier OSNR in moderate-distance systems (e.g. data centre interconnects, and metro/regional networks), and hence the achievable overall bitrate. For microcombs, the per-line power is typically lower than benchtop systems, requiring amplification which reduces the OSNR. This can be mitigated, for example, by filtering the comb lines with a narrow band cavity filter having an FSR matched to the comb line spacing, [73,91] since the optical noise is broad whereas the individual comb lines are very narrow-band. Using a simple passive microring filter has improved the OSNR by > 10 dB, meaning that the current state of the art OSNRs for microcombs of 30-40 dB can be pushed to 40-50 dB – on par or better than benchtop comb sources. This conceptually simple filtering process to distil the noise of many comb lines simultaneously could represent the final key to enable benchtop performance from microcombs.

One rapidly developing field for optical communications is free-space, for both terrestrial and satellite communications [92, 93]. For satellites in particular, reducing the size, weight and power of the system is critical, and so microcombs are ideal. Directional free-space optical communications are being investigated for high data rate links between earth and low earth orbit satellites [94, 95] and for future missions to the moon by NASA. [96, 97] The current record data rate for earth-orbit communications is on the order of a single legacy coherent fibre communication channel (i.e. 1-200 Gb/s) [96] and this can only be maintained for a short time before thermal management issues force an extended shutdown. Providing higher data rates continuously between locations needs significant reduction in power consumption while maintaining a compact footprint and weight. Microcombs have supported transmission through the atmosphere, with 100 lines at 10 Gb/s per channel using differential detection phase shift keying modulation protocol, yielding an aggregate rate of 1 Tb/s over 10 km [92]. While this distance is much shorter than the 100's of km to orbit, it nonetheless produces a similar degree of turbulence. This protocol has the advantage of being broadly insensitive to the resulting phase fluctuations but is difficult to scale to high efficiencies. Adaptive optics has been used with commercial coherent transponders to enable over 13 Tb/s [93] over 10 km at sea level, although this did not use a comb source. Microcombs [98-125] have already had a major impact on microwave photonics, neuromorphic processing and signal processing [126-161]. This emerging field of communications offers yet further new and exciting opportunities for microcombs to have a significant impact.

In summary, microcombs have made tremendous strides in research demonstrations of ultrahigh bandwidth massively parallel optical communications in the past 10 years. They now have achieved many milestone requirements including turnkey self-starting, self-recovery, efficient operation and high performance in terms of comb-line power and OSNR. Much progress has been made in achieving full integration with different platforms, thus foreseeing the day when fully integrated microcomb based transceivers will be available, resulting in extremely low SWaP and cost. Their commercialization for real-world systems has already been taken up by several new startup companies, arguably heralding a new era for optical frequency microcomb chips, barely 15 years after their invention.

## VI. Conclusions

We review recent progress on ultrahigh bandwidth optical fiber communications based on integrated optical frequency comb technologies, or integrated Kerr microcombs. Remarkable progress has been achieved in the past few years with numerous record-breaking demonstrations. Progress in microcomb technologies has arguably made them competitive with benchtop systems but with the added enormous advantages in cost and size, weight and power. And yet we are still in the early days of this exciting field – it has barely been half a dozen years since the first demonstration of coherent microcomb based ultrahigh bandwidth communications. While microcombs have notably matured they are still an ongoing area of research. Reducing the size and power consumption is key, as is improving stand-alone long-term reliability and ease of generation, or turn-key operation. Microcombs have been one of the hottest research fields for the past 10 years and for good reason. They offer the promise of the full capability of laboratory-based frequency combs but in an ultra-compact, reliable, efficient, cost effective integrated platform. Optical communications have arguably been one of the most successful applications of microcombs and it is exciting and inspiring to think that this may be one of the first commercial industries where they will have a profound and lasting impact.

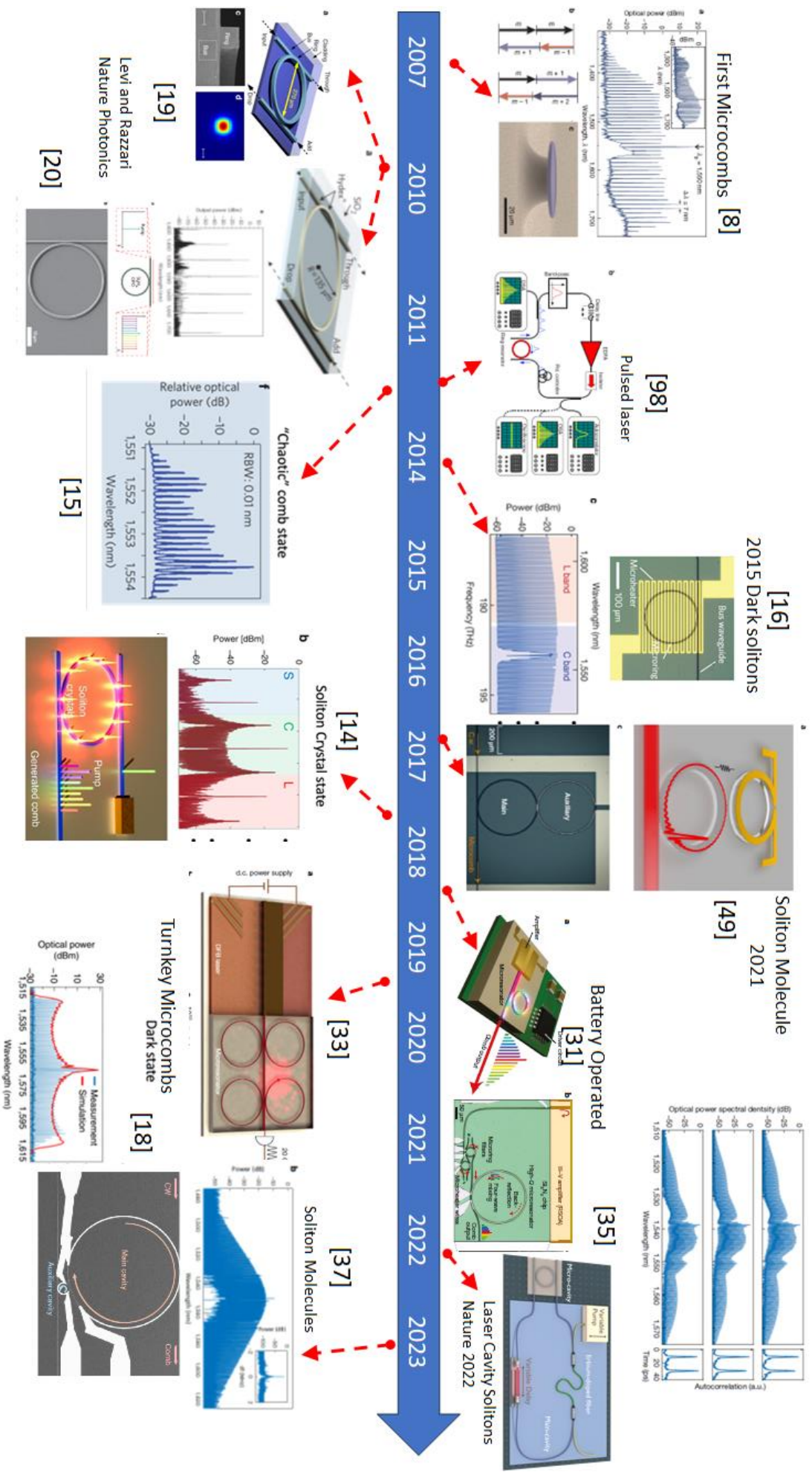

**Figure 1.** Evolution of Optical Microcombs.

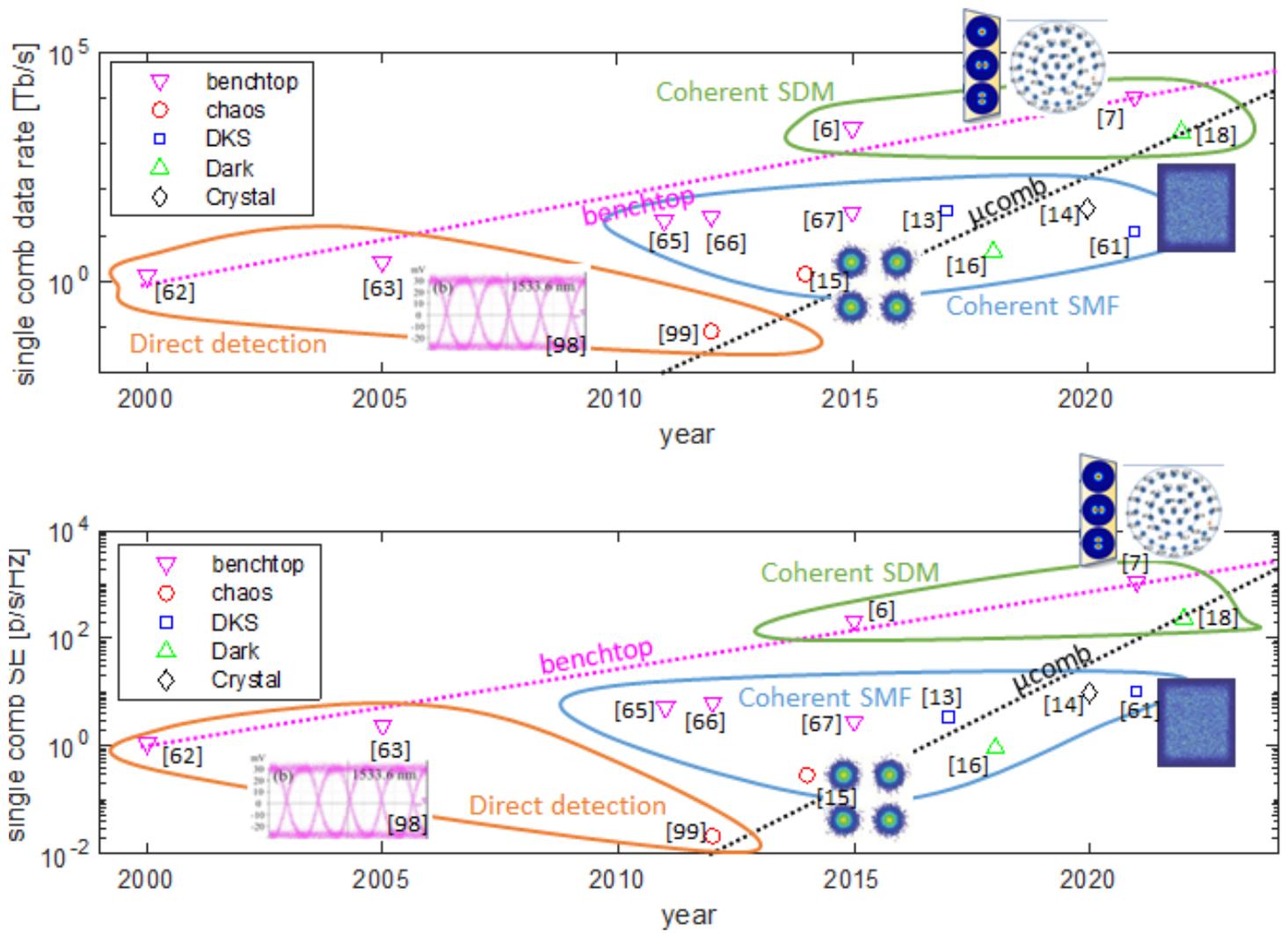

**Figure 2.** Evolution of Optical Communications with Benchtop combs and microcombs. Top: bandwidth versus time and Bottom: spectral density versus time.

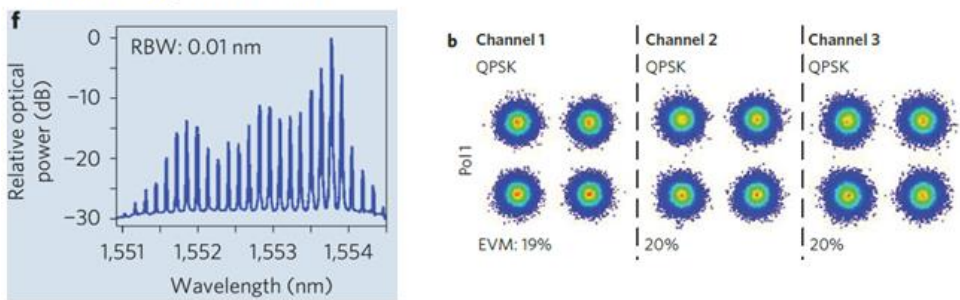
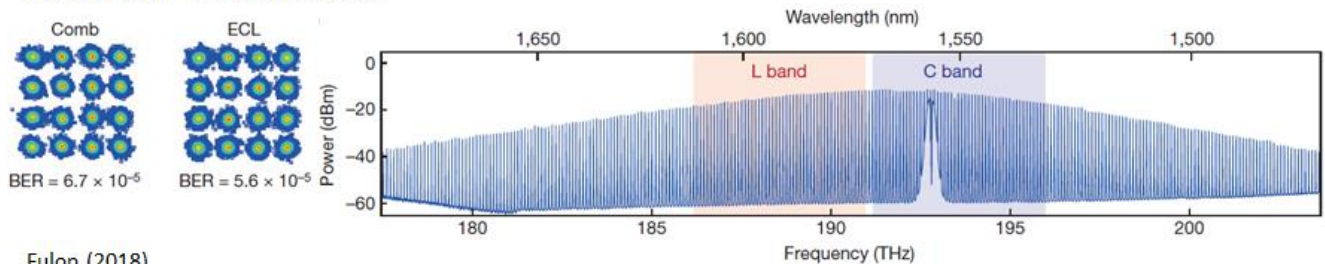
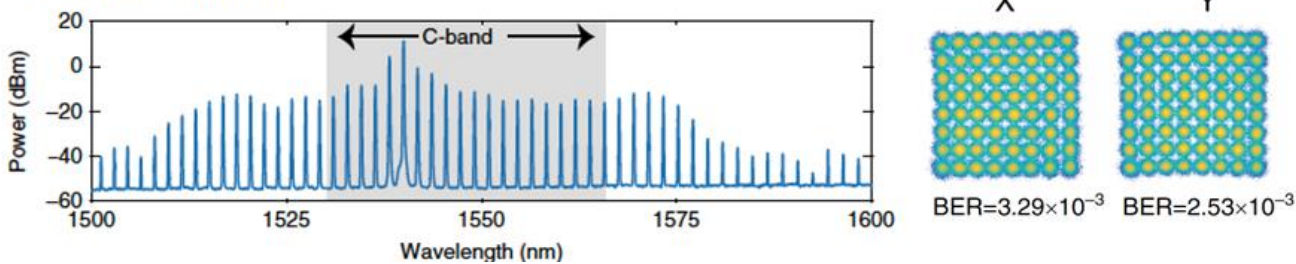

**Figure 3.** Recent demonstrations of optical communications system based on microcombs. Top Ref[15]. Middle Ref [13], Bottom from Ref[16]

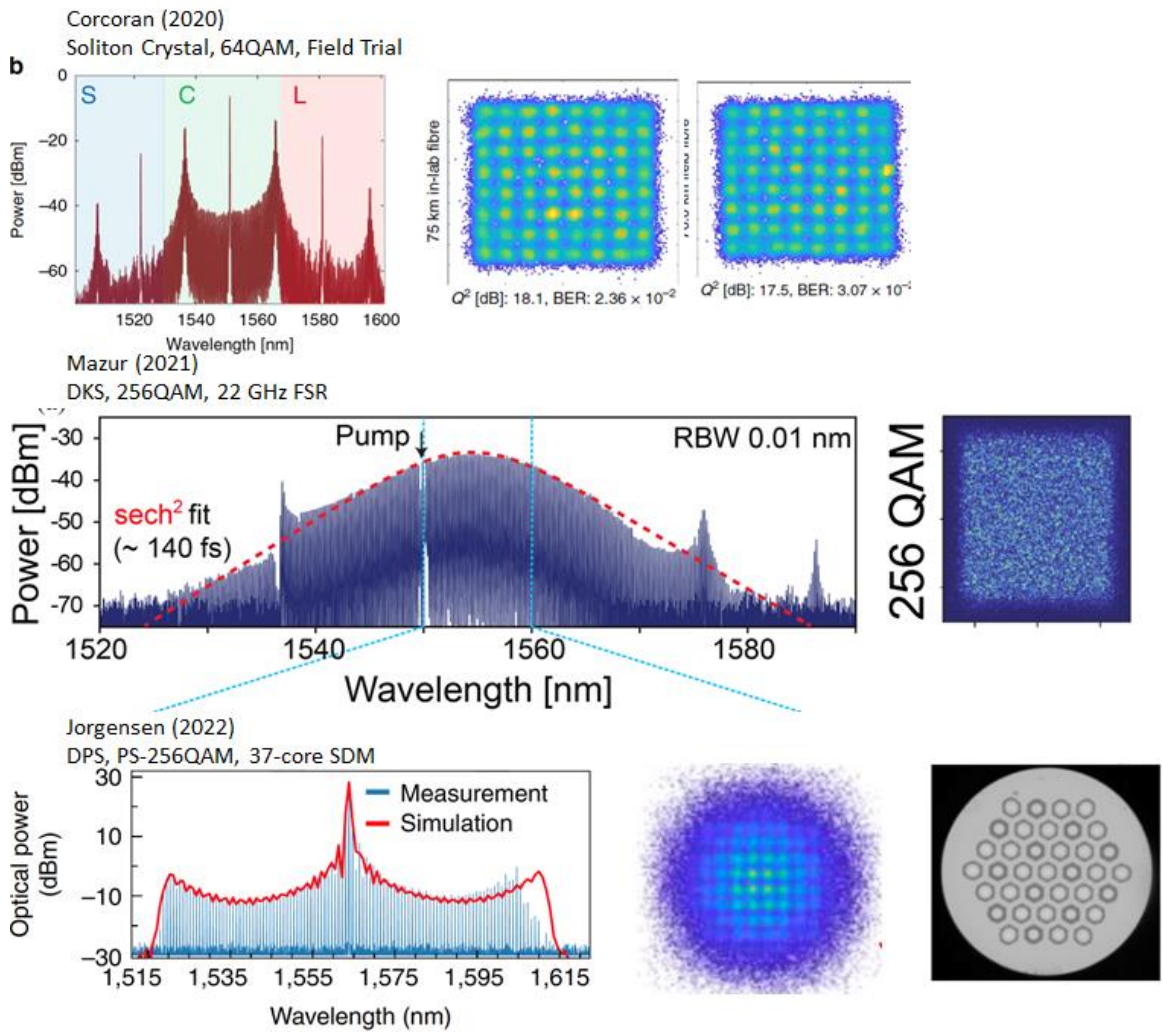

**Figure 4.** Recent demonstrations of optical communications system based on microcombs. Top from reference [14]. Middle from Ref. [61], Bottom from Ref [18].

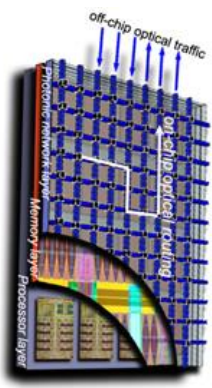
On chip optical interconnects

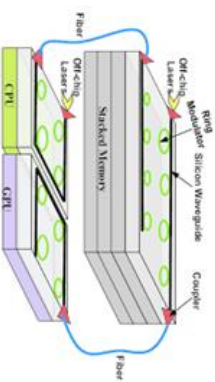
Chip-chip

100μm — 1cm — 1m

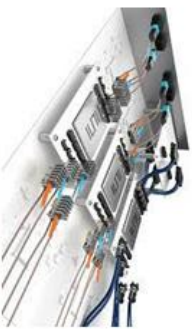
Board-board

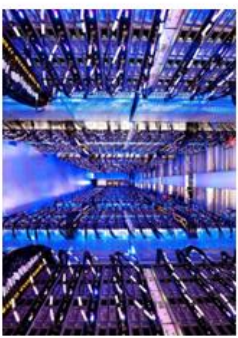
Intra-Data Centre

100m

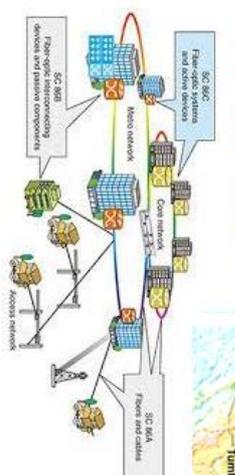
Metro Networks

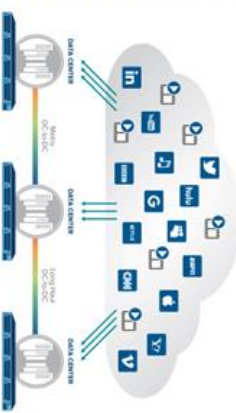
Inter-Data Centre

1-10km

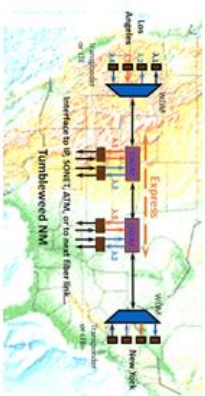
Longhaul Terrestrial

100km

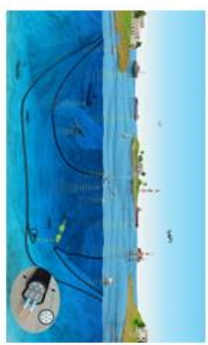
Undersea

1,000km

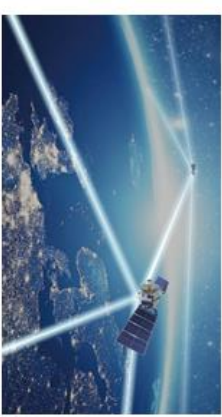
Satellite to Satellite

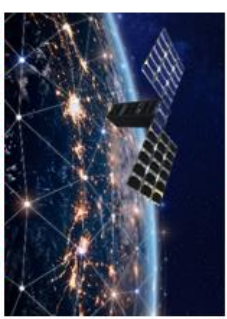
Satellite to/from Earth

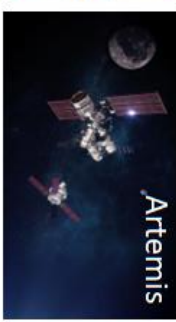
Lunar / Deep Space
Artemis

10,000km

**Figure 5.** Potential applications of optical microcombs in real-world systems.

# Table 1

| paper | year | rate (per Tx comb) [Tb/s] | Bandwidth (& FSR) | comb type |
|---|---|---|---|---|
| P-H. Wang [99] | 2012 | 0.08 | 3.75T (596G) | Chaos |
| Pfieffle [15] | 2014 | 1.44 | 5T (25G) | Chaos |
| Marin-Polomo [13] | 2017 | 34.6 | 9.9T (95.8G) | DKS |
| Fulop [16] | 2018 | 4.4 | 4.6T (230G) | Dark pulse |
| Corcoran [14] | 2020 | 39 | 3.95T (48.9G) | Soliton Crystal |
| Mazur [61] | 2021 | 12 | 1.14T (22G) | DKS |
| Jørgensen [18] | 2022 | 1840 | 7.8T (105→35G) | Dark pulse |